# Modelling and predicting labor force productivity

Ivan O. Kitov, Oleg I. Kitov

*Abstract*
Labor productivity in Turkey, Spain, Belgium, Austria, Switzerland, and New Zealand has been analyzed and modeled. These counties extend the previously analyzed set of the US, UK, Japan, France, Italy, and Canada. Modelling is based on the link between the rate of labor participation and real GDP per capita. New results validate the link and allow predicting a drop in productivity by 2010 in almost all studied countries.

**Introduction**

We continue reporting results of the study devoted to the driving force behind labor productivity, *P*. In [1], we presented a nonlinear and lagged link between productivity and real GDP per capita, *G*, in the United States, Japan, the United Kingdom, France, Italy, and Canada. These countries are the largest developed economies in the world.

We defined two components of the growth rate of *G* – a trend (also potential or neutral growth) and fluctuations. The trend component is proportional to the reciprocal value of the attained level of *G*, *A/G*, where *A* is an empirical country-dependent constant. The fluctuations are driven by the change in some specific age population. By subtracting *A/G* from *dG/G* one obtains the driving force of the change in productivity, as well as of the rate of labor force participation [2].

In developed countries, population estimates for the specific age are not available or too poor for quantitative analysis. However, for the modelling of the changes in productivity one can use estimates of real GDP per capita instead of the population estimates. This paper extends the set of studied countries and presents the link between *P* and *G* in Turkey, Spain, Belgium, Austria, Switzerland, and New Zealand. Moreover, the lag of productivity behind the change in real GDP allows predicting the former at various time horizons.

The remainder of the paper is organized as follows. Section 1 presents the model developed in [1,2] as a set of quantitative relationships between labor productivity, labor force participation rate, the growth rate of real GDP per capita. In Section 2, we continue testing these relationships against actual data and present some predictions of the future evolution of productivity in all six studied countries.

**1. The model**

For the estimation of labor productivity one needs to know total output (GDP) and the level of employment, *E* (*P=GDP/E*), or total number of working hours, *H* (*P=GDP/H*). In the first approximation and for the purposes of our modelling, we neglect the difference between the employment and the level of labor force because the number of unemployed is only a small portion of the labor force. There is no principal difficulty, however, in the subtraction of the unemployment, which is completely defined by the level of labor force with possible complication in some countries induced by time lags [3,4]. The number of working hours is an independent measure of the workforce. Employed people do not have the same amount of working hours. Therefore, the number of working hours may change without any change in the level of employment and vice versa. In this study, the estimates associated with *H* are not used.

Individual productivity varies in a wide range in developed economies. In order to obtain a hypothetical true value of average labor productivity one needs to sum up individual productivity of each and every employed person with corresponding working time. This definition allows a proper correction when one unit of labor is added or subtracted and distinguishes between two states with the same employment and hours worked but with different productivity. Hence, both standard definitions are slightly biased and represent approximations to the true productivity. Due to the absence of the true estimates of labor productivity and related uncertainty in the approximating definitions we do not put severe constraints on the precision in our modelling and seek only for a visual fit between observed and predicted estimates.

In this study, we use the estimates of productivity and real GDP per capita reported by the Conference Board (http://www.conference-board.org/economics/database.cfm). Recently, we developed a model [2] describing the evolution of labor force participation rate, $LFP$, in developed countries as a function of a single defining variable – real GDP per capita. Natural fluctuations in real economic growth unambiguously lead to relevant changes in labor force participation rate as expressed by the following relationship:

$$\{B_1 dLFP(t)/LFP(t) + C_1\} \exp\{\alpha_1 [LFP(t) - LFP(t_0)]/LFP(t_0) =$$
$$= \int \{dG(t-T))/G(t-T) – A_1/G(t-T)\}dt \qquad (1)$$

where $B_1$ and $C_1$ are empirical (country-specific) calibration constants, $\alpha_1$ is empirical (also country-specific) exponent, $t_0$ is the start year of modelling, $T$ is the time lag, and $dt=t_2-t_1$, $t_1$ and $t_2$ are the start and the end time of the time period for the integration of $g(t) = dG(t-T))/G(t-T) – A_1/G(t-T)$ (one year in our model). Term $A_1/G(t-T)$, where $A_1$ is an empirical constant, represents the evolution of economic trend [4]. The exponential term defines the change in sensitivity to $G$ due to the deviation of the $LFP$ from its initial value $LFP(t_0)$. Relationship (1) fully determines the behavior of $LFP$ when $G$ is an exogenous variable.

It follows from (1) that labor productivity can be represented as a function of $LFP$ and $G$, $P \sim G \cdot Np/Np \cdot LFP = G/LFP$, where $Np$ is the working age population. Hence, $P$ is a function of $G$ only. Therefore, the growth rate of labor productivity can be represented using several independent variables. Because the change in productivity is synchronized with that in $G$ and labor force participation, first useful form mimics (1):

$$dP(t)/P(t) = \{B_2 dLFP(t)/LFP(t) + C_2\} \cdot \exp\{\alpha_1 [LFP(t) - LFP(t_0)]/LFP(t_0)\} \qquad (1')$$

where $B_2$ and $C_2$ are empirical calibration constants. Inherently, the participation rate is not the driving force of productivity, but (1′) demonstrates an important feature of the link between $P$ and $LFP$ – the same change in the participation rate may result in different changes in the productivity depending on the level of the $LFP$.

In order to obtain a simple functional dependence between $P$ and $G$ one can use two alternative forms of (1), as proposed in [1]:

$$\{B_3 dLFP(t)/LFP(t) + C_3\} \exp\{\alpha_2[LFP(t) - LFP(t_0)]/LFP(t_0)\} = N_s(t-T)$$

$$dP(t)/P(t) = B_4 N_s(t-T) + C_4 \quad (2)$$

where $N_s$ is the number of $S$-year-olds, i.e. in the specific age population, $B_3,..., C_4$ are empirical constant different from $B_2$, $C_2$, and $\alpha_2=\alpha_1$. In this representation, we use our finding that the evolution of real GDP per capita is driven by the change rate of the number of $S$-year-olds. Relationship (2) links $dP/P$ and $N_s$ directly.

The following relationship defines $dP/P$ as a nonlinear function of $G$ only:

$$N(t_2) = N(t_1) \cdot \{ 2[dG(t_2-T)/G(t_2-T) - A_2/G(t_2-T)] + 1\} \quad (3)$$

$$dP(t_2)/P(t_2) = N(t_2-T)/B + C \quad (4)$$

where $N(t)$ is the (formally defined) specific age population, as obtained using $A_2$ instead of $A_1$; $B$ and $C$ are empirical constants. Relationship (3) defines the evolution of some specific age population, which is different from actual one.

1. **Productivity prediction**

In this Section, we use relationships (3) and (4) for the prediction of labor productivity in Turkey, Spain, Belgium, Austria, Switzerland, and New Zealand. These countries extend the previous set of the largest developed economies in the world.

Figure 1 presents the growth rate of productivity in Turkey - observed and predicted one. The observed values are presented by open circles and the predicted ones with solid diamonds. Because of strong fluctuations in original time series the observed curve is smoothed with a 3-year moving average, MA(3). Real GDP per capita is obtained from the Conference Board data base [2008] in 1990 GK dollars. Productivity estimates ($ per working person per year) are also

taken from the same database. The predicted rate is obtained using (3) and (4) from real GDP per capita.

The observed rate has been varying from ($dP/P$=) 0.07 $y^{-1}$ to -0.02 $y^{-1}$ in 1980 and near 2000. Real economic growth has been oscillating around its very low potential rate defined by $A_2$=$105. Both productivity curves in Figure 1 are well synchronized with all major peaks matched. Linear regression analysis gives $R^2$=0.51 for the period between 1966 and 2006. All in all, the predicted curve is in excellent agreement with the observed one and this observation confirms our previous results reported in [1].

An outstanding and expected feature of the predicted curve is that the change in real GDP leads the growth in productivity by 2 years. In Figure 1, the predicted curve is shifted by 2 years back ($T$=2 year) in order to synchronize it with the measured one. This lead allows prediction of the future evolution of productivity in Turkey at a two-year horizon. After 2005, the productivity has been suffering a dramatic fall that will continue into 2010. Such a dynamic change during a short period will be used to validate relationships (3) and (4) with the above parameters.

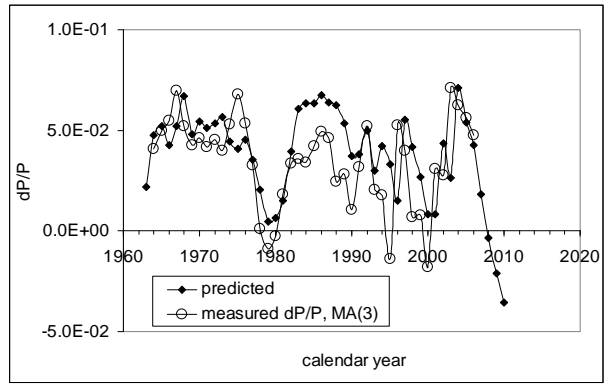

Figure 1. Observed and predicted (from real GDP pee capita) change rate of productivity in Turkey. The observed curve is represented by MA(3) of the original version. Model parameters are as follows: $A_2$=$105, N(1959)=1450000, B=-6000000, C=0.24, T=2 year.

Figure 2 depicts observed and predicted productivity for Spain. These curves are similar to those for France [1] and are also in an excellent agreement: $R^2$=0.9. Such high correlation is likely a biased result because both series are non-stationary. The change in productivity in Spain varies from 0.1 $y^{-1}$ in the 1960s to -0.01 $y^{-1}$ in the 2000s. Hence, real economic growth has been far below its potential rate ($A_2$=$175) since 1960s. The current rate of productivity growth is negative and one should not expect any break in the declining trend. Surprisingly, there is no time lag of the productivity behind the change in real GDP, $T$=0.

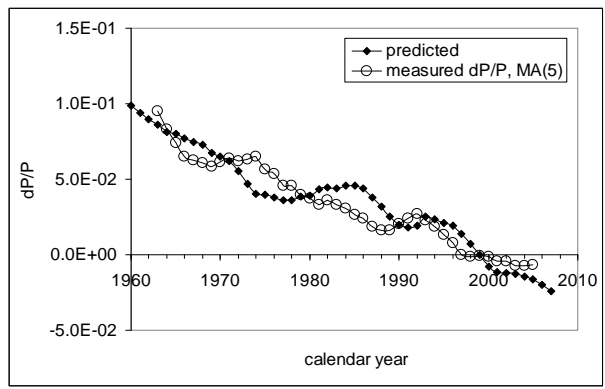

Figure 2. Observed and predicted (from real GDP pee capita) change rate of productivity in Spain. The observed curve is represented by MA(5) of original version. Model parameters are as follows: $A_2$=$175, $N$(1959)=1050000, $B$=-3000000, $C$=0.13, $T$=0 year.

The case of Belgium may be considered as a standard one. Figure 3 displays measured and predicted rate of productivity growth. The curves are very close with $R^2$=0.78 for the period between 1967 and 2007. For Belgium, the range of productivity change is smaller than in many developed countries: from 0.05 $y^{-1}$ in the 1970s to 0.01 $y^{-1}$ in the 2000s. The current rate of productivity growth is also close to 0.01 $y^{-1}$. However, Belgium is characterized by a 5-year lag of the productivity. This value is not abnormal but is close to the largest lags. The rate of neutral (or potential) growth is not the highest one as defined by $A_2$=$280.

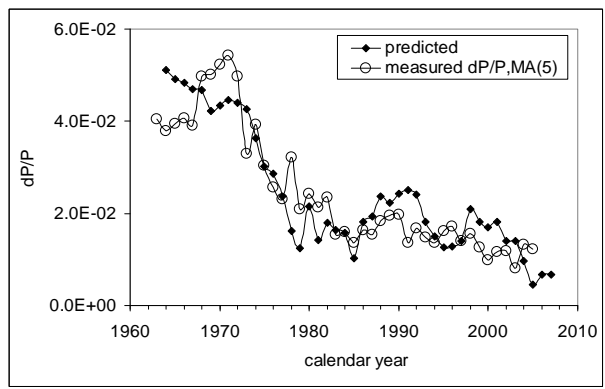

Figure 3. Observed and predicted (from real GDP pee capita) change rate of productivity in Belgium. The observed curve is represented by MA(5) of original version. Model parameters are as follows: $A_2$=$280, $N$(1959)=150000, $B$=-1900000, $C$=0.13, $T$=5 year.

The evolution of productivity in Austria is presented in Figure 4. Currently, labor productivity in the Austrian economy evolves at a very low rate near 0.01 $y^{-1}$. This is not a new situation – after 1975 the rate has been hovering between 0.01 $y^{-1}$ and 0.02 $y^{-1}$. An outstanding feature is the rate of potential growth defined by $A_2$=$335, almost the largest among developed countries. This rate is three times higher than in Turkey and twice as big as in Spain, when referred to the same level of real GDP per capita. This demonstrates a remarkable efficiency of the Austrian economy.

As in many developed countries, productivity in Austria lags behind the change in real GDP by 3 years. This lag allows predicting a sudden drop in the growth rate of productivity to negative figures in 2010. Considering high correlation ($R^2$=0.8) between the observed and predicted curves since 1963 the drop in the growth rate is practically inevitable. At the same time, the predicted drop will serve as a validation of the model.

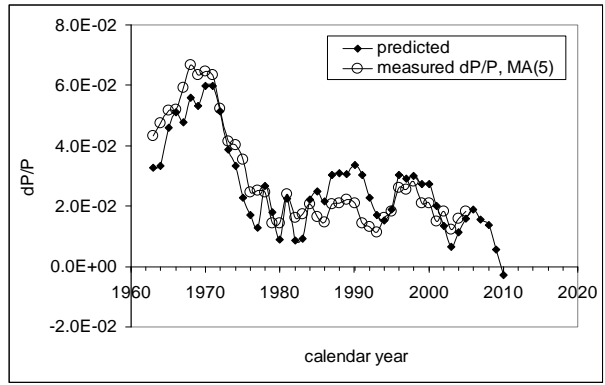

Figure 4. Observed and predicted (from real GDP pee capita) change rate of productivity in Austria. The observed curve is represented by MA(5) of original version. Model parameters are as follows: $A_2$=$335, $N(1959)$=100000, $B$=-500000, $C$=0.243, $T$=3 year.

Switzerland and New Zealand are presented in Figures 5 and 6. They are similar in terms of time lag: $T$=4 years in both countries, and the rate of neutral growth defined by $A_2$=$175 and $A_2$=$170, respectively. In both countries, relatively accurate prediction from $G$ is possible only after ~1975. The discrepancy before 1970 is not well explained and might be linked to revisions to employment and real economic growth definitions, and/or measurement errors. In both countries, the rate of productivity growth will approach the zero line by 2010.

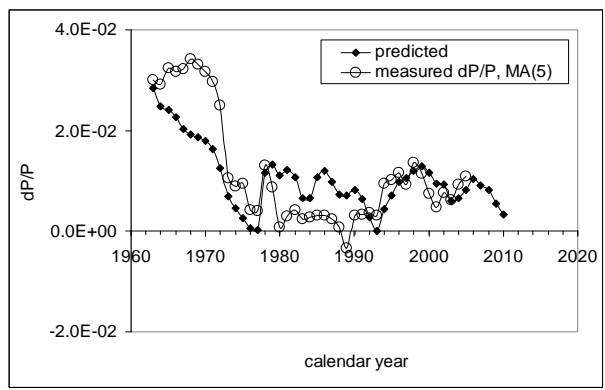

Figure 5. Observed and predicted (from real GDP pee capita) change rate of productivity in Switzerland. The observed curve is represented by MA(5) of original version. Model parameters are as follows: $A_2$=$175, $N(1959)$=200000, $B$=-4500000, $C$=0.076, $C$=0.243, $T$=4 year.

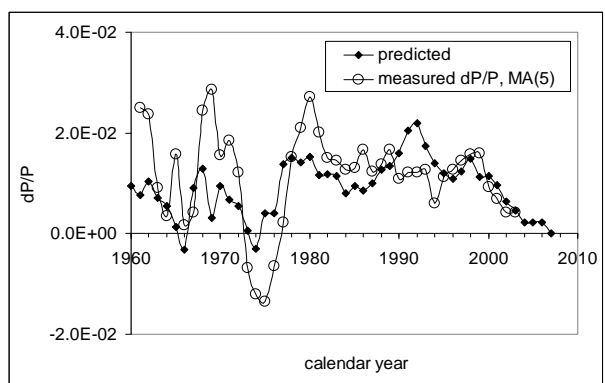

Figure 6. Observed and predicted (from real GDP pee capita) change rate of productivity in New Zealand. The observed curve is represented by MA(5) of original version. Model parameters are as follows: $A_2$=$170, $N(1959)$=40000, $B$=-550000, $C$=0.076, $C$=0.243, $T$=4 year.

**Conclusion**

We have successfully modelled labor productivity in Turkey, Spain, Belgium, Austria, Switzerland, and New Zealand. These six countries extend the previously modelled set consisting of the largest economies. Therefore, our concept is valid: labor productivity is a secondary macroeconomic variable because it is completely defined by the growth in real GDP per capita relative to its neutral rate, $A_2/G$. Since real economic growth depends only on the evolution of specific age population, one has to care about demographic processes in order to control labor productivity and stable economic growth.